\documentclass{article}
\usepackage{spconf,amsmath,graphicx}
\usepackage{siunitx}
\usepackage{url}
\newcommand{\me}{\mathrm{e}}
\usepackage{tabularx}
\usepackage[table]{xcolor}
\usepackage{booktabs}
\usepackage{multirow}
\newcolumntype{C}{>{\Centering\arraybackslash\hspace{0pt}}X}
\usepackage{hyperref}
\usepackage{fancyhdr}

\title{Characterizing the temporal dynamics of universal speech representations for generalizable deepfake detection}
\name{Yi Zhu, Saurabh Powar\sthanks{Saurabh is a student at Veermata Jijabai Technological Institute and assisted with the work during his internship at INRS-EMT.}, and Tiago H. Falk
}
\address{Institut national de la recherche scientifique, INRS-EMT, University of Québec, Montréal, Canada}
\fancypagestyle{FirstPage}{
\lfoot{\copyright 2020 IEEE. Personal use of this material is permitted. Permission from IEEE must be obtained for all other uses, in any current or future media, including reprinting/republishing this material for advertising or promotional purposes, creating new collective works, for resale or redistribution to servers or lists, or reuse of any copyrighted component of this work in other works.} 
}

\begin{document}
%\ninept
%
\maketitle

\begin{abstract}
Existing deepfake speech detection systems lack generalizability to unseen attacks (i.e., samples generated by generative algorithms not seen during training). Recent studies have explored the use of universal speech representations to tackle this issue and have obtained inspiring results. These works, however, have focused on innovating downstream classifiers while leaving the representation itself untouched. In this study, we argue that characterizing the long-term temporal dynamics of these representations is crucial for generalizability and propose a new method to assess representation dynamics. Indeed, we show that different generative models generate similar representation dynamics patterns with our proposed method. Experiments on the ASVspoof 2019 and 2021 datasets validate the benefits of the proposed method to detect deepfakes from methods unseen during training, significantly improving on several benchmark methods. 
\end{abstract}
%the time-varying nature of universal representations makes them sub-optimal for generalizing to unseen attacks, since the captured temporal nuances might be specific to a certain type of generative model. We here propose to emphasize the long-term dynamics of speech by converting the time-varying universal representations to a temporally static dynamic representation using an innovated modulation transformation block. 

\section{Introduction}
\label{sec:intro}
\thispagestyle{FirstPage}
Deepfake (DF) speech refers to speech signals generated either by text-to-speech (TTS) synthesis and/or voice conversion systems (VC). With the burgeoning interests in generative models, an increasing number of tools have become available to craft deepfake speech, of which the naturalness is almost indistinguishable from the genuine ones~\cite{petra2023}. Like a double-edged sword, such technique can pose serious dangers if used improperly, such as gaining access to a voice-guarded system or cloning users' voice to commit fraud and scam~\cite{firc2022dawn}.

To tackle these potential threats, numerous efforts have been made including the curation of DF speech datasets and challenges (e.g., ASVspoof 2019 and 2021 ~\cite{todisco2019asvspoof, yamagishi2021asvspoof}), as well as the development of DF speech detection models~\cite{todisco2019asvspoof, yamagishi2021asvspoof}. The majority of DF speech detection models rely on supervised training with no pre-training involved, such as the linear frequency cepstral coefficients (LFCC) based systems~\cite{chen2020generalization} and the RawNet variants~\cite{tak2021end}, which were used as baseline systems in multiple DF detection challenges~\cite{todisco2019asvspoof, yamagishi2021asvspoof}. However, such training strategy leads to low generalizability to samples manipulated by unseen algorithms (i.e., unseen attacks). For example, high equal error rates (EER) were reported in the ASVspoof 2021 DF track~\cite{liu2023asvspoof}, of which the evaluation set contains multiple unseen DF generation algorithms.

More recently, studies have explored the use of universal speech representations learned from self-supervised learning for DF speech detection. Representative include wav2vec2~\cite{baevski2020wav2vec}, Hubert~\cite{hsu2021hubert}, and wavLM~\cite{chen2022wavlm}. These representations are commonly pre-trained on massive amounts of data, hence encompassing richer information compared to conventional speech features. For example, several works have explored the use of wav2vec as the front-end encoder appended with different downstream classification layers for DF prediction~\cite{wang2021investigating, wang2022fully, tak2022automatic, lv2022fake}; some of them already showed improved generalizability to unseen attacks~\cite{wang2022fully, tak2022automatic}. 

Most of these works, however, focused on innovating downstream classifier architectures, leaving the universal representations untouched. However, as these representations are generated at high temporal resolution (e.g., \SI{50}{Hz} for wavLM embeddings~\cite{chen2022wavlm}), we argue that though this helps to capture the short-time changes (e.g., linguistic content), the long-term dynamics of the representations are not optimally represented. Long-term dynamics may capture insights about syllabic rate, articulatory movement, and/or other vocal attributes (e.g., hoarseness), thus should be crucial for detecting DF speech. More importantly, while certain abnormalities may hide in the short-time changes, for example, the mispronunciation of a syllable or a glitch artifact, such nuances may not generalize well to unseen generative algorithms. As such, characterizing long-term temporal dynamics should make systems more generalizable.

In this study, we show the importance of long-term temporal dynamics of the universal speech representations to detect DF speech, especially out-of-domain samples. In particular, we characterize the dynamics from two universal representations, namely wav2vec2 and wavLM. To characterize the long-term dynamics, we propose a method similar to the speech modulation spectrum, widely used in speech applications (e.g., \cite{ding2017temporal, tiwari2022modulation}), where we take a short-time Fourier transform (STFT) across the time dimension of the raw feature $\times$ time representation. By changing the window size of the STFT, we can characterize the long-term dynamics of the representation. Experimental results show consistent improvement in generalizability to out-of-domain samples with both wav2vec2 and wavLM encoders. The proposed method can be easily integrated with most mainstream classifiers and representations, thus can be applied to new methods. Codes used in the experiments herein have been made publicly available at \href{https://github.com/zhu00121/Universal-representation-dynamics-of-deepfake-speech}{our Github repository}.

\vspace{-2mm}
\section{Deepfake Speech Detection Systems}
\label{sec:system}
%This section describes DF speech detection systems experimented in this study, including two baseline systems and two universal representation based systems (wav2vec2 and wavLM) together with their variants.

\subsection{Baseline Systems}
The first baseline system uses LFCC features (i.e., energies and 1st- and 2nd-order deltas) extracted from 19 spectral bands, which serve as input to a Gaussian mixture model (GMM) classifier. This method was shown to outperform several large deep learning based models in the ASVspoof challenges~\cite{todisco2019asvspoof, yamagishi2021asvspoof}. The second baseline takes the raw audio waveform as input and employs the RawNet2 as the downstream classifier~\cite{yamagishi2021asvspoof}. RawNet2 is an end-to-end network tailored specifically for anti-spoofing, comprising sinc filters, convolutional blocks, and a GRU layer~\cite{tak2021end}. 

\subsection{Universal Representation Encoders}
We employed the wav2vec2-960h and wavLM-base-plus as the speech encoders~\cite{baevski2020wav2vec, chen2022wavlm}. These encoders share similar model architecture and audio processing pipelines. The raw waveform (sampled at \SI{16}{kHz}) is sent as input, which firstly passes through a convolutional neural network (CNN) block. This helps to compress the highly complex waveform into a more compact representation (feature $\times$ time). The stride of each CNN kernel is \SI{20}{ms}, which leaves the output from the CNN with a temporal resolution of \SI{50}{Hz}. This is then fed into the cascaded transformer layers, where the temporal resolution remains unchanged. Both encoders were pre-trained in a self-supervised manner on a large amount of data, where the output from intermediate transformer layers have shown to be useful for different speech tasks~\cite{baevski2020wav2vec, chen2022wavlm}. The pre-trained models were downloaded from HuggingFace.

\subsection{Characterizing Long-Term Temporal Dynamics of Unsupervised Representations}
\begin{figure}
    \centering
    \includegraphics[width=0.8\linewidth]{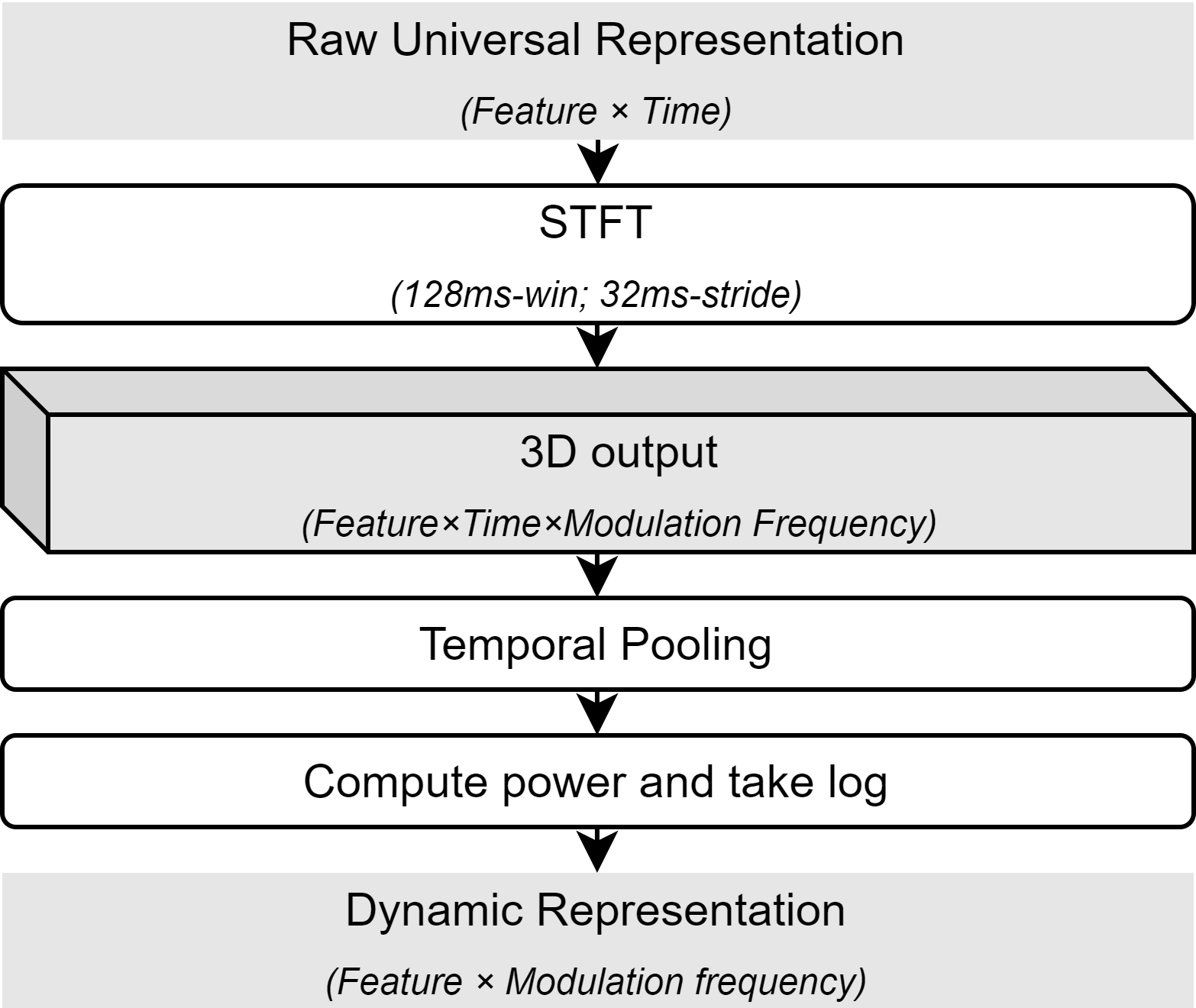}
    \caption{Computation pipeline of the proposed modulation transformation block.}
    \label{fig:mtb}
    \vspace{-3mm}
\end{figure}

While the universal representations are assumed to already encode different speech aspects, we argue that the time-varying nature of such representations makes them sub-optimal for generalizing to out-of-domain samples. We propose to overcome this limitation by integrating a so-called modulation transformation block (MTB), akin to the modulation spectral processing methods widely used in speech processing and computed from spectrograms. The computation steps involved in a MTB are depicted in Fig.~\ref{fig:mtb}. 

The input to the MTB module is the aggregation of 2-dimensional outputs from all transformer layers in the pre-trained encoders. Previous studies have shown that the importance of different layers vary for different tasks~\cite{chen2022wavlm}, however, it is not clear which layers are most crucial for discriminating DF speech. Hence, we applied 1-dimensional learnable weights to all layers, which were initialized with same values and updated during training; finally, we used the weighted sum of all layer outputs. The resultant representation shares the same shape as the output of each transformer layer (768 $\times$ time), where 768 corresponds to the feature dimensionality of the wav2vec2 and wavLM representations. 

Next, a short-time Fourier transform (STFT) is applied to each feature channel across time. Here, we use a fixed window length of \SI{128}{ms} and hop length of \SI{32}{ms} to capture long-term temporal dynamics. This value was based on pilot experiments where window sizes of \SI{128}{ms}, \SI{512}{ms}, and \SI{1024}{ms} were explored. The STFT decomposes the temporal changes within each \SI{128}{ms}-window into different ``modulation frequency bins'', thus characterizing the long-term representation dynamics. By sliding windows along the time axis, a 3D tensor is generated (feature $\times$ modulation frequency $\times$ time). As we aim to emphasize the overall characteristic of the entire utterance, an average temporal pooling is then applied across the time axis. Modulation energies are then computed and the $\log$ operation is taken for numeric stability. In general, the MTB module maps the 2D universal speech representation (feature $\times$ time) into another 2D representation (feature $\times$ modulation frequency) which characterizes the long-term temporal dynamic pattern of the entire utterance.
 
\subsection{Downstream Classification Layers}
Previous studies have shown improved generalizability by appending complex classification layers subsequently to the encoders~\cite{tak2022automatic, lv2022fake}. Such an approach increases model size and decreases system interpretability and explainability. Here, as we are interested in understanding the usefulness of the proposed method itself, we leave optimization of classification layers for future investigation. We compared two types of representations, one being the original (raw) universal representation (with weighted sum) and the other the proposed temporal dynamics representation processed by the MTB. Both are pooled into a 1-dimensional vector with 768 feature values. Two fully-connected (FC) layers with a dropout layer are used to map the representations into a final decision. For fair comparisons, the number of hidden neurons in the FC layers was kept the same for both representations.

\vspace{-2mm}
\section{Experimental Setup}
\label{sec:setup}
\subsection{Datasets}
{\bf ASVspoof 2019 (LA track):} The Logical Access (LA) track of the ASVspoof 2019 challenge is derived from the multi-speaker VCTK corpus~\cite{todisco2019asvspoof}. The DF utterances in the evaluation set were generated using 17 different text-to-speech (TTS) and VC algorithms, among which six were seen in the training and validation sets while the remaining 11 were unseen attacks. \\
{\bf ASVspoof 2021 (DF track):} A DF track was added to the 2021 ASVspoof challenge~\cite{yamagishi2021asvspoof}. While the training and validation data remained the same as those used in the ASVspoof 2019, the evaluation set was increased to 600K utterances generated by more than 100 different spoofing algorithms. Furthermore, the data conditions and audio compression techniques of the evaluation data also differed from that of the training and validation set, hence posing a more strict task compared to the ASVspoof 2019 challenge.

\subsection{Metrics and Statistical Analysis}
Similar to the two ASVspoof challenges~\cite{todisco2019asvspoof, yamagishi2021asvspoof}, the EER was used as the main metric for model evaluation, which is the rate where the false acceptance rate is equal to the false rejection rate. The lower the EER value is, the more accurate the detection system is. Meanwhile, we also report the F1 score, which is a commonly used metric for unbalanced binary classification tasks. To measure if two compared models performed significantly differently in EERs, we employed the method from \cite{bengio2004statistical}, which conducts a pair-wise statistical analysis.

\subsection{Training and Inference Details}
With the two baseline systems, we followed the same pipelines as described in the ASVspoof challenge code repositories with the model hyper-parameters unchanged. Interested readers are encouraged to refer to \cite{todisco2019asvspoof} and \cite{yamagishi2021asvspoof} for more details. For a fair comparison between different variants of the universal representation based systems, we adopted the exact same training strategies as well as model hyper-parameters.

The universal representation encoders were frozen during training, only the learnable weights assigned to the encoder transformer layers and downstream classification layers were updated. This resulted in a total of fewer than \SI{5}{million} parameters for all models. The binary cross entropy (BCE) loss was used, where the weights assigned to genuine samples were set to 10 to tackle the class imbalance. For computational efficiency, the batch size was set to 1 and the maximal training epochs were set to 20. The learning rate was decreased linearly from $1\me^{-4}$ to $1\me^{-5}$. To avoid over-fitting, the training process was stopped when the EER values did not decrease for three consecutive epochs. The FC layer dropout value was set to 0.25. Model training and inference were conducted on the Compute Canada cluster~\cite{baldwin2012compute} with four V100l GPUs. Average computing time was around \SI{8}{h} per model (training+inference).

\vspace{-2mm}
\section{Results and Discussion}
\subsection{Model Performance}
We first compare the model performance obtained using the ASVspoof 2021 evaluation set (Table~\ref{tab:p1}), where the majority of the test samples were generated by unseen algorithms. Among all tested systems, the proposed wavLM temporal dynamics representation based system is shown to be the top-performing one with an EER of 9.87\% and F1 of 0.403, surpassing the ASVspoof 2021 DF track top-1 result. The wav2vec2 based systems, on the other hand, achieved only baseline-level performance. When comparing the proposed representation versus the original universal (raw) representation, a consistent statistically significant improvement is seen with both wav2vec2 and wavLM after applying the proposed modulation transformation. Interestingly, while the proposed representation performs better with the evaluation data, the raw universal representations achieve lower EER scores on the validation set, indicating potential model over-fitting to seen attacks. Such finding corroborates with our initial hypothesis that although universal representations encompass rich temporal details, which can be important for detecting DF speech, part of the information learned is unique to each generative algorithm and may not generalize well to unseen attacks.

We further report the performance achieved with the ASVspoof 2019 evaluation set, where 6 out of the 17 algorithms were seen during training (Table~\ref{tab:p2}). In accordance with the ASVspoof 2021 results, wavLM is shown to be a stronger encoder than wav2vec2. Meanwhile, it can be noticed that both encoders perform markedly better on the ASVspoof 2019 data, suggesting that detecting seen attacks is a much simpler task than detecting unseen attacks. Similar to the ASVspoof 2021 validation results, the top-performer here is shown to be the raw wavLM representation based system. To further quantify the representation generalizability, we calculated the gap between the EERs achieved with ASVspoof 2019 and ASVspoof 2021 data. A significantly lower gap is seen with the proposed wavLM temporal dynamics representation compared to the raw one. Together, these findings demonstrate that while universal representations can be directly used to differentiate between genuine and DF speech, generalizability can be further improved by characterizing long-term temporal dynamics. 

\begin{table}[]
\caption{Performance achieved using ASVspoof 2021 DF track data. Statistical significance was measured between EERs of each pair of universal representations (raw and proposed) with the significantly better score highlighted in grey.}
\centering
\begin{tabularx}{\linewidth}{ccccc}
\toprule
\multirow{2}{*}{Model} & \multicolumn{2}{c}{Valid} & \multicolumn{2}{c}{Eval}\\
\cmidrule{2-3} 
\cmidrule{4-5}
& EER (\%) & F1 & EER (\%) & F1 \\
\midrule
LFCC+GMM & 3.76 & .834 & 25.56 & .197\\
RawNet2 & 3.07 & .858 & 22.38 & .213\\
Top-1 result~\cite{yamagishi2021asvspoof} &  - & - & 15.64 & -\\
\midrule
Wav2vec2-raw & \cellcolor{lightgray}4.35 & .819 & 28.00 & .160\\
Wav2vec2-proposed & 9.47 & .662 & \cellcolor{lightgray}27.10 & .166\\
\midrule
WavLM-raw & .08 & .996 & 10.89 & .380 \\
WavLM-proposed & .78 & .963 & \cellcolor{lightgray} 9.87 &  .403 \\
\bottomrule
\end{tabularx}
\vspace{-3mm}
\label{tab:p1}
\end{table}

\begin{table}[]
\caption{Performance achieved with ASVspoof 2019 LA track data. EER gap suggests the difference in EERs obtained with ASVspoof 2021 and 2019 evaluation data. Statistical significance was measured between EERs and the EER gap of each pair of universal representations.}
\centering
\begin{tabularx}{\linewidth}{cccc}
\toprule
\multirow{2}{*}{Model} & \multicolumn{2}{c}{ASVspoof 2019 Eval} & \multirow{2}{*}{EER gap}\\
\cmidrule{2-3} 
& EER (\%) & F1 & \\
\midrule
LFCC+GMM & 25.30 & .387 & .26\\
RawNet2 & 7.46 & .745 & 14.72\\
\midrule
Wav2vec2-raw & \cellcolor{lightgray} 13.20 & .576 & 12.44\\
Wav2vec2-proposed & 16.40 & .513 & 11.60\\
\midrule
WavLM-raw & \cellcolor{lightgray} .72 &  .966 & 10.22\\
WavLM-proposed & 2.47 & .891 & \cellcolor{lightgray} 7.40\\
\bottomrule
\end{tabularx}
\label{tab:p2}
\vspace{-2mm}
\end{table}

% \vspace{-2mm}
\subsection{Visualizing Temporal Dynamic Representations}
\begin{figure}
    \centering
    \includegraphics[width=\linewidth]{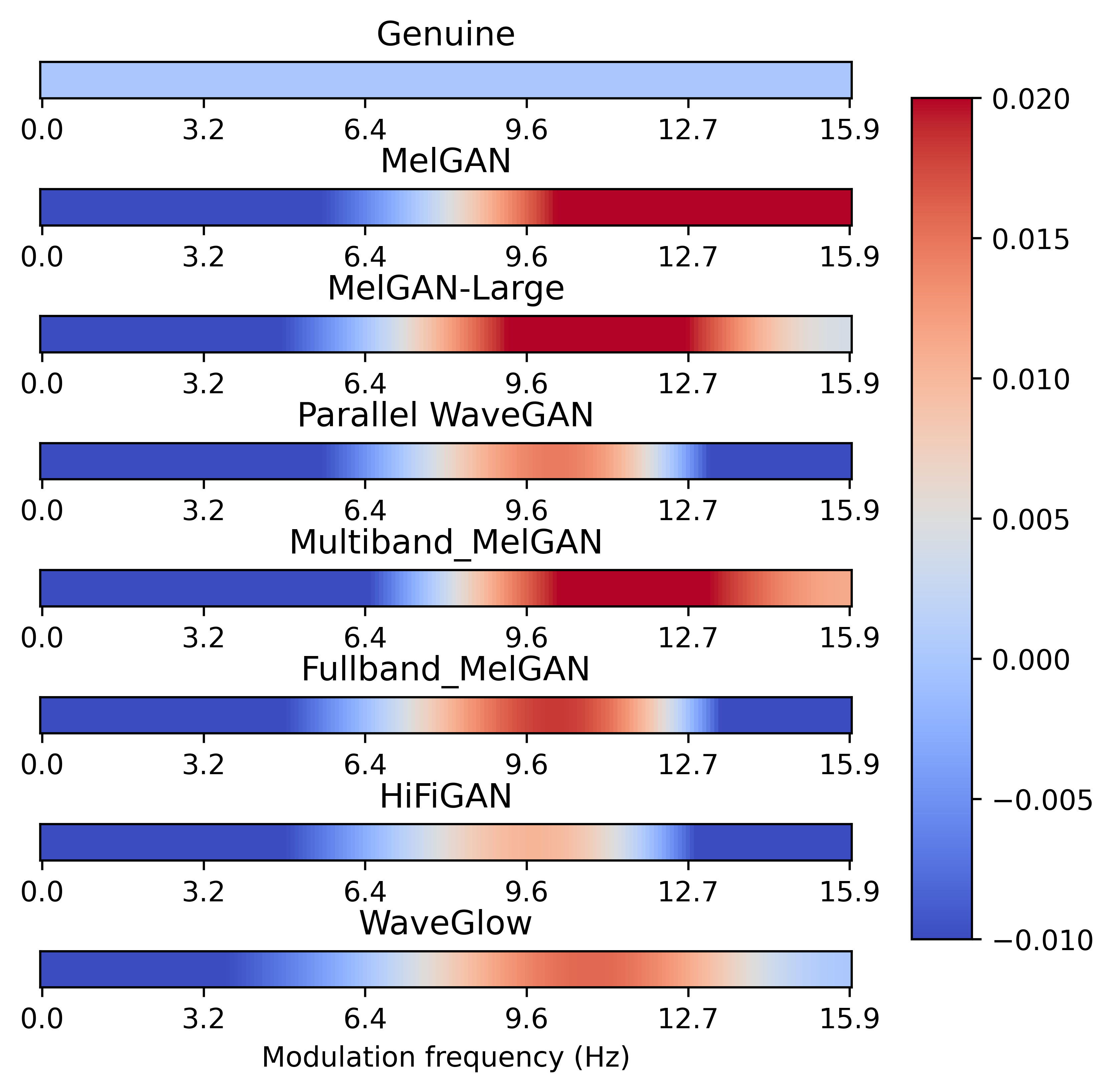}
    \caption{Visualization of 1-dimensional compressed version of the proposed representation computed for a genuine utterance and deepfake versions generated by seven different algorithms. The genuine pattern is subtracted from all for better comparison.}
    \label{fig:dynamics}
\vspace{-3mm}
\end{figure}

Next, we visualized our proposed representation extracted from different types of DF speech. Since no text ground-truth nor speaker info was provided with the utterances from ASVspoof~\cite{yamagishi2021asvspoof}, we selected one exemplary utterance from the LJspeech corpus~\cite{ljspeech17} and seven different deepfake versions from the WaveFake corpus~\cite{frank2021wavefake} which contains DF speech generated using samples from LJspeech. All eight utterances share the exact same speech content and correspond to the same speaker. We computed the proposed 2d temporal dynamic representations from all utterances and averaged over the feature axis to highlight the modulation spectral patterns. For better visualization, the spectral pattern of the genuine one is subtracted from all eight utterances. The resultant plots are shown in Fig.~\ref{fig:dynamics}. Regions with higher spectral energies are indicated by warmer colors, suggesting more discrimination compared to the genuine utterance. Since all other vocal attributes are controlled, the difference seen here is likely caused by the generation process. A spectral peak between \SI{7}{} to \SI{12}{Hz} can be observed across all seven algorithms, while some also demonstrate peaks in the higher modulation frequency range (e.g., higher energies above \SI{12}{Hz} for MelGAN and HiFiGAN). Such consistency in the modulation spectral pattern corroborates with the improved generalizability obtained by applying the modulation transformation. A possible explanation is that though unseen attacks are generated by models different from those used in the training set, the abnormalities in long-term temporal dynamics can be similar, leading to a consistent modulation spectral pattern that can be used for speech deepfake detection for unseen attacks.

\vspace{-2mm}
\section{Conclusion}
In this study, we showed the importance of characterizing the long-term temporal dynamics of unsupervised speech representations to increase the generalizability of deepfake speech detection. The representation was shown to convey similar patterns across seven different deepfake detection methods, resulting in significantly better detection results and improved generalizability to unseen attacks. 

% % Below is an example of how to insert images. Delete the ``\vspace'' line,
% % uncomment the preceding line ``\centerline...'' and replace ``imageX.ps''
% % with a suitable PostScript file name.
% % -------------------------------------------------------------------------
% \begin{figure}[htb]

% \begin{minipage}[b]{1.0\linewidth}
%   \centering
%   \centerline{\includegraphics[width=8.5cm]{image1}}
% %  \vspace{2.0cm}
%   \centerline{(a) Result 1}\medskip
% \end{minipage}
% %
% \begin{minipage}[b]{.48\linewidth}
%   \centering
%   \centerline{\includegraphics[width=4.0cm]{image3}}
% %  \vspace{1.5cm}
%   \centerline{(b) Results 3}\medskip
% \end{minipage}
% \hfill
% \begin{minipage}[b]{0.48\linewidth}
%   \centering
%   \centerline{\includegraphics[width=4.0cm]{image4}}
% %  \vspace{1.5cm}
%   \centerline{(c) Result 4}\medskip
% \end{minipage}
% %
% \caption{Example of placing a figure with experimental results.}
% \label{fig:res}
% %
% \end{figure}

\vfill\pagebreak

\bibliographystyle{IEEEbib}
\bibliography{strings,main}

\end{document}